\documentstyle[preprint,prb,aps]{revtex}
\begin{document}
\draft

\title{Specific heat and validity of quasiparticle approximation\\
in the half-filled Landau level}

\author{Yong Baek Kim}
\address{Bell Laboratories, 600 Mountain Avenue, Murray Hill, NJ 07974}
\author{Patrick A. Lee}
\address{Department of Physics, Massachusetts Institute of Technology,
Cambridge, MA 02139}

\maketitle

\begin{abstract}
We calculate the specific heat of composite fermion
system in the half-filled Landau level.
Two different methods are used to examine validity of
the quasiparticle approximation when the two-body
interaction is given by $V(q) = V_0 / q^{2-\eta}$ 
($1 \le \eta \le 2$).
The singular part of the specific heat is 
calculated from the free energy of the gauge field,
which is compared with the specific heat calculated
from the quasiparticle approximation via the 
singular self-energy correction due to the gauge 
field fluctuations.
It turns out that two results are in general 
different and they coincide only for the case
of the Coulomb interaction ($\eta = 1$).
This result supports the fact that 
the quasiparticle approximation is valid only
for the case of the Coulomb interaction.
It is emphasized that this result is obtained by
looking at a gauge-invariant quantity -- the specific heat. 
\end{abstract}
\pacs{Pacs numbers: 73.40.Hm, 71.10.Pm, 11.15.-q}

The appearance of the unexpected metallic state at
the filling fraction $\nu =1/2$ in the fractional quantum
Hall regime has stimulated 
a lot of activities in theory
\cite{Jain,Lopez,HLR,Kalmeyer,Kim1,Millis,Kim2,Stern,Kim3}
\cite{Stamp,Polchinski,Gan,Kwon,Nayak,He,Simon}
and experiment
\cite{Jiang,Egap,leadley,mass,manoharan,Willet1,Kang,Goldman,Willet2}.
The similarity between the phenomena around $\nu = 1/2$ 
in high magnetic fields and those of electrons in weak 
magnetic fields is successfully explained by the concept
of composite fermions\cite{Jain}.
A composite fermion is obtained by attaching even number
of flux quanta to an electron\cite{Jain,Lopez,HLR}. 
In particular, in the half-filled Landau
level, one attaches two flux quanta to an electron to
make a composite fermion. 
At the mean field level, if one takes the average of
the statistical magnetic field coming from the attached
magnetic flux, the composite fermions at $\nu = 1/2$ feel 
zero magnetic field due to the cancellation between the 
external magnetic field and the averaged statistical 
magnetic field\cite{HLR}. 
As a result, the mean field ground state
is a filled Fermi-sea with a well-defined Fermi wave
vector $k_F$\cite{HLR}. There exist several experiments which 
demonstrate the existence of a well-defined Fermi wave 
vector at $\nu = 1/2$\cite{Willet1,Kang,Goldman,Willet2}.

Note that the fluctuations of the statistical magnetic field 
correspond to the density fluctuations due to the fact that
the statistical flux quanta are attached to each electron. 
Therefore, in the mean field approximation, the strong density 
fluctuations are ignored.
This implies that the fluctuations about the mean field state
can be very important and are basically gauge field 
fluctuations. 
The above arguments also suggest that the effects of the
two-particle interaction $v({\bf q}) = V_0 / q^{2-\eta}$
($1 \le \eta \le 2$) on the gauge field fluctuations should
be examined carefully. In fact, the gauge field 
fluctuations become more singular as the interaction range
becomes shorter (larger $\eta$) because
the longer range interaction (smaller $\eta$) suppresses 
the density fluctuations more effectively\cite{HLR,Kim1}.
Thus, it is important to examine the stability of the 
mean field Fermi-liquid state against the gauge field 
fluctuations. 

In order to study systematically the effects of the gauge 
field fluctuations, Halperin, Lee, and Read (HLR) used the 
fermionic Chern-Simons gauge theory to develop a theory of
the metallic state at $\nu = 1/2$\cite{HLR}.
In the random phase approximation which becomes exact in
the large $N$ limit ($N$ is the number of species 
of fermions), the most singular correction to the self-energy
comes from the transverse part of the gauge field, which is 
given by\cite{HLR}
\begin{equation}
D_{11} ({\bf q},\nu) = {1 \over -i \gamma \nu / q + 
\chi q^{\eta}} \ ,
\label{gaugefield}
\end{equation}
where $\gamma = 2 n_e / k_F$, $\chi = {1 \over 24 \pi m} + 
{V_0 \over (2 \pi {\tilde \phi})^2}$ for $\eta = 2$, and
$\chi = {V_0 \over (2 \pi {\tilde \phi})^2}$ for 
$1 \le \eta < 2$.
The correction to the retarded self-energy in the lowest order 
perturbation theory is given by\cite{HLR,Kim1,Nagaosa}, 
for $1 < \eta \le 2$,
\begin{equation}
\Sigma^R ({\bf k},\omega) = 
\left [ {\rm Re} \lambda + i \ {\rm Im} \lambda \ 
{\rm sgn} (\omega) \right ] 
|\omega|^{2 \over 1 + \eta} \ ,
\end{equation}
where
\begin{equation}
\lambda = 
{v_F \ e^{i {\pi \over 2} 
\left ( {\eta - 1 \over \eta + 1} \right )} \over 4 \pi \  
{\rm sin} \left ( {2 \pi \over 1 + \eta} \right )
\gamma^{\eta - 1 \over \eta + 1} \chi^{2 \over 1 + \eta}} \ .
\end{equation}
For $\eta = 1$, we have
\begin{equation}
\Sigma^R ({\bf k},\omega) = {v_F \over 4 \pi^2 \chi} \ |\omega|
\left [ {\rm ln} \left ( {4 k^2_F \chi \over \gamma |\omega|} \right )
+ i {\pi \over 2} {\rm sgn} (\omega) \right ] \ .  
\end{equation}
Therefore, the usual Landau criterion for the quasiparticle is
violated in the case of $1 < \eta \le 2$ and the case of
$\eta = 1$ shows the marginal Fermi liquid behavior.

In Ref.~\cite{HLR}, an attempt was made to construct a renormalized 
quasiparticle theory using the above singular self-energy
correction even though there is no well-defined quasiparticle
in the usual sense of Landau-Fermi-liquid theory. 
It was assumed that there exits a well defined
Fermi wave vector $k_F = \sqrt{4 \pi n_e}$ and, 
for $|{\bf k}| \approx k_F$, there exit
quasiparticle excitations with the energy spectrum\cite{HLR},
for $1 < \eta \le 2$, 
\begin{equation}
E ({\bf k}) = 
\left ( |\xi_{\bf k}| / {\rm Re} \lambda 
\right )^{1 + \eta \over 2} {\rm sgn} (\xi_{\bf k}) \ ,
\label{short}
\end{equation}
where $\xi_{\bf k} = {k^2 \over 2m} - \mu$.     
For $\eta = 1$, the quasiparticle spectrum becomes
\begin{equation}
E ({\bf k}) = {4 \pi^2 \chi \over v_F} \ \xi_{\bf k} \ 
{\rm ln} \left ( {k^2_F v_F \over \pi^2 
\gamma |\xi_{\bf k}|} \right ) \ .
\label{long}
\end{equation}
Note that these quasiparticle spectra are obtained 
from the real part of the retarded self-energy.

Recently we derived and studied the quantum Boltzmann 
equation (QBE) of the composite fermions interacting with 
the gauge field\cite{Kim3}. It was emphasized that a generalized 
QBE can be constructed even though there is no 
well defined Landau-quasiparticle\cite{Kim3}.
In particular, the collision integral of the QBE becomes
relatively unimportant only when the two-particle interaction
is given by the Coulomb interaction ($\eta = 1$)\cite{Kim3}.
In this case, the QBE is equivalent to the usual QBE of
Landau-Fermi-liquid theory.
For the case of Coulomb interaction, Stern and 
Halperin\cite{Stern}
showed that the usual Landau-Fermi-liquid theory can
be successfully applied, which is consistent with the
construction of the QBE mentioned above.
Also Kwon, Houghton, and Maston\cite{Kwon} argued that 
the higher
dimensional bosonization can be consistently performed
only for the case of the Coulomb interaction ($\eta = 1$). 
This result may imply that the case of the Coulomb 
interaction can be described by the usual 
Landau-Fermi-liquid theory.

In this paper, we want to examine the range of the 
applicability of the quasiparticle approximation in
relation to the Landau-Fermi-liquid theory.
In particular, we address this issue by
looking at a gauge-invariant physical quantity -- specific heat.
According to the standard procedure, the total specific heat of 
the system can be calculated
as the sum of the contributions from the free fermion and
the gauge field\cite{HLR}. 
The free fermion part gives linear temperature dependence
while the gauge field part gives more singular non-linear
temperature dependence in the low temperature limit.
In the quasiparticle approximation, the singular correction
to the specific heat could be considered as coming from the 
singular mass renormalization of the fermions.
In fact, both of the methods give the same temperature 
dependence for the singular part of the specific heat.
The question is whether they are exactly the same in the
low temperature limit.     
Thus, as a direct test of the quasiparticle 
approximation, we calculate the most singular 
part of the specific heat using two different methods.
One is to use the dispersion relation of the 
renormalized quasiparticle in the quasiparticle
approximation. The other way is to calculate the
specific heat from the free energy of the gauge field.
The latter method is supposed to give the 
correct answer because the latter one is
gauge-invariant while the former is not. 
It turns out that the two results are in general 
different and they agree with each other only for
the case of the Coulomb interaction.
This result supports the fact that the quasiparticle 
approximation can be safely used only for the case
of the Coulomb interaction. We want to emphasize that
this result is obtained by looking at a 
gauge-invariant quantity -- the specific heat.

First, let us evaluate the specific heat using the
quasiparticle approximation. 
Using the dispersion relation given by Eqs.~\ref{short}
and \ref{long}, the expectation value of the energy 
can be obtained as
\begin{equation}
\langle E \rangle = \sum_{\bf k} E ({\bf k}) f ({\bf k}) \ ,
\end{equation}
where $f ({\bf k}) = 1 / (e^{E ({\bf k})/T} + 1)$.
The specific heat can be evaluated from $C_{\rm qp} = 
{\partial \langle E \rangle \over \partial T}$:
\begin{equation}
{\partial \langle E \rangle \over \partial T} =
{m \over 4 \pi} \int^{\infty}_0 d \xi_{\bf k} \ 
{E^2({\bf k}) / T^2 \over [{\rm cosh} (E({\bf k})/2T)]^2} \ .
\end{equation} 
For $1 < \eta \le 2$, it is given by
\begin{equation}
C_{\rm qp} (T) = \alpha_{\rm qp}(\eta) \ T^{2 \over 1 + \eta} \ ,
\end{equation}
where
\begin{equation}
\alpha_{\rm qp} (\eta) =
{1-2^{-{2 \over 1 + \eta}} \over 2 \pi (1 + \eta)} \ 
{1 \over {\rm sin} 
\left ( {\pi \over 2}{\eta - 1 \over \eta + 1} \right )} \
\Gamma \left ( {2 \eta + 4 \over \eta + 1} \right )
\zeta \left ( {\eta + 3 \over \eta + 1} \right ) 
\left ( {\gamma \over \chi} \right )^{2 \over 1 + \eta} \ .
\end{equation}
In the case of $\eta = 1$, we have
\begin{equation}
C_{\rm qp} (T) = {1 \over 12} \ {\gamma \over \chi} \ T \ 
{\rm ln} \left ( 4 k^2_F \chi \over \gamma T \right ) \ .
\label{qp}
\end{equation}

Now we calculate the specific heat from the free
energy of the gauge field:
\begin{equation}
F_{\rm g} = \int^{\infty}_{-\infty} {d \nu \over 2 \pi}
\int {d^2 q \over (2 \pi)^2} \ {1 \over e^{\nu / T} - 1} \ 
{\rm arctan} \left ( 
{{\rm Im} D^{-1}_{11} \over {\rm Re} D^{-1}_{11}}
\right ) \ .
\end{equation}
The contribution of the gauge field to the entropy
can be obtained as
\begin{equation}
S_{\rm g} = - {\partial F_{\rm g} \over \partial T}
= \int^{\infty}_{-\infty} {d \nu \over 2 \pi}
\int^{\infty}_0 {q dq \over 2 \pi} \ {\nu / T^2 \over 
(e^{\nu / 2T} - e^{- \nu / 2T})^2} \
{\rm arctan} \left ( {\gamma \over \chi}
{\nu \over q^{1 + \eta}} \right ) \ .
\end{equation}
Thus, for $1 < \eta \le 2$, the specific heat $C_{\rm g} = 
T {\partial S_{\rm g} \over \partial T}$ is given by
\begin{equation}
C_{\rm g} (T) = \alpha_{\rm g} (\eta) \ T^{2 \over 1 + \eta} \ ,
\end{equation}
where
\begin{equation} 
\alpha_{\rm g} (\eta) =
{1 \over 4 \pi (1 + \eta)} \ 
{1 \over {\rm sin} 
\left ( {\pi \over 2}{\eta - 1 \over \eta + 1} \right )} \
\Gamma \left ( {2 \eta + 4 \over \eta + 1} \right )
\zeta \left ( {\eta + 3 \over \eta + 1} \right ) 
\left ( {\gamma \over \chi} \right )^{2 \over 1 + \eta} \ .
\end{equation}
In the case of $\eta = 1$, the specific heat becomes
the same as Eq.~\ref{qp}.

Note that $C_{\rm g} (T)$ is the correct answer for the
singular part of the specific heat because it is calculated
in a gauge-invariant way.
Note also that $C_{\rm g} (T)$ agrees with
the result $C_{\rm qp} (T)$ of the quasiparticle 
approximation only for the case of $\eta = 1$.
The underlying reason for this behavior is as follows.
In the quasiparticle approximation, only the real part
of the self-energy is used to get the spectrum of the
quasiparticle. However, both of the real and imaginary
parts of the self-energy contribute to the specific heat,
{\it i.e.}, one has to also take into account the 
information about the imaginary part of the self-energy.
The sum of these contributions are basically the same as the
specific heat calculated from the free energy of the
gauge field\cite{Kim2}.
For the case of the Coulomb interaction, the imaginary
part of the self-energy is logarimically smaller than
the real part of the self-energy so that the real
part of the self-energy is sufficient to get the 
correct answer for the most singular part of the 
specific heat.
This result supports the previous conclusion that the 
quasiparticle approximation can
be justified only when the two-particle interaction is
given by the Coulomb interaction
($\eta = 1$).   

We would like to thank B.~I.~Halperin, F.~D.~M.~Haldane and 
X.-G.~Wen for helpful discussions about this and related issues in 
relation to the half-filled Landau level.
PAL is supported by NSF grant No. DMR-9523361.

\end{document}